\documentclass[iop,numberedappendix]{emulateapj}

\usepackage{graphicx,amsmath,amssymb,natbib}
\usepackage{here}
\usepackage{epsf}
\usepackage{epstopdf}
\usepackage{xspace}
\usepackage{hyperref}

\usepackage[normalem]{ulem} % for regular text; use \sout{}
\usepackage{cancel} % for equations; use \cancel{}

\usepackage{color}

\newcommand{\um}{\ensuremath{\mu\rm{m}}\xspace}
\newcommand{\kms}{\ensuremath{\rm{km\,s}^{-1}}\xspace}
\newcommand{\alphaco}{\ensuremath{\alpha_{\rm{CO}}}\xspace}

\newcommand{\uJy}{\ensuremath{\mu\rm{Jy}}\xspace}

\newcommand{\lir}{\ensuremath{L_{\rm{IR}}}\xspace}
\newcommand{\Tdust}{\ensuremath{T_{\rm{dust}}}\xspace}

\newcommand{\Mstar}{\ensuremath{M_{\rm{star}}}\xspace}
\newcommand{\Mgas}{\ensuremath{M_{\rm{gas}}}\xspace}
\newcommand{\Mdyn}{\ensuremath{M_{\rm{dyn}}}\xspace}
\newcommand{\msol}{\ensuremath{\rm{M}_\odot}\xspace}

\newcommand{\fgas}{\ensuremath{f_{\rm{gas}}}\xspace}
\newcommand{\tdep}{\ensuremath{t_{\rm{dep}}}\xspace}
\newcommand{\lprime}{\ensuremath{\rm{L}_{\rm{CO(1-0)}}'}\xspace}

\newcommand{\nii}{[N{\scriptsize II}]\xspace}
\newcommand{\halpha}{\ensuremath{\rm{H}\alpha}\xspace}

\newcommand{\gn}{GOODS-N~774\xspace}
\newcommand{\ctt}{COSMOS~22995\xspace}
\newcommand{\cts}{COSMOS~27289\xspace}

\begin{document}

\title{Low Gas Fractions Connect Compact Star-Forming Galaxies \\ to their \MakeLowercase{z} $\sim$ 2 Quiescent Descendants}

\def\Arizona{1}
\def\UMass{2}
\author{Justin S. Spilker$^{\Arizona}$}
\author{Rachel Bezanson$^{\Arizona,\dagger}$}
\author{Daniel P. Marrone$^{\Arizona}$}
\author{Benjamin J. Weiner$^{\Arizona}$}
\author{Katherine E. Whitaker$^{\UMass,\dagger}$}
\author{Christina C. Williams$^{\Arizona}$}
\altaffiltext{\Arizona}{Steward Observatory, University of Arizona, 933 North Cherry Avenue, Tucson, AZ 85721, USA; \href{mailto:jspilker@as.arizona.edu}{jspilker@as.arizona.edu}}
\altaffiltext{\UMass}{Department of Astronomy, University of Massachusetts, Amherst, MA 01003, USA}
\altaffiltext{\textdagger}{Hubble Fellow}

\shortauthors{J. S. Spilker, et al.}
\shorttitle{Low Gas Fractions in Compact Star-Forming Galaxies}

%%%%%%%%%%%%%%%%%%%%%%%%%%%%%%%%%%%%%%%%%%%%%%%%%%%%%%%%%%%%%%%%%%%%%%%%%%%%%%%%%%%%%
%%%%%%%%%%%%%%%%%%%%%%%%%%%%%%%%%%%% ABSTRACT %%%%%%%%%%%%%%%%%%%%%%%%%%%%%%%%%%%%%%%
%%%%%%%%%%%%%%%%%%%%%%%%%%%%%%%%%%%%%%%%%%%%%%%%%%%%%%%%%%%%%%%%%%%%%%%%%%%%%%%%%%%%%
\begin{abstract}

Early quiescent galaxies at $z\sim2$ are known to be remarkably compact compared to their nearby counterparts. Possible progenitors of these systems include galaxies that are structurally similar, but are still rapidly forming stars. Here, we present Karl G. Jansky Very Large Array (VLA) observations of the CO(1--0) line towards three such compact, star-forming galaxies at $z\sim2.3$, significantly detecting one. The VLA observations indicate baryonic gas fractions $\gtrsim$5 times lower and gas depletion times $\gtrsim$10 times shorter than normal, extended massive star-forming galaxies at these redshifts. At their current star formation rates, all three objects will deplete their gas reservoirs within 100\,Myr. These objects are among the most gas-poor objects observed at $z>2$, and are outliers from standard gas scaling relations, a result which remains true regardless of assumptions about the CO-H$_2$ conversion factor. Our observations are consistent with the idea that compact, star-forming galaxies are in a rapid state of transition to quiescence in tandem with the build-up of the $z\sim2$ quenched population. In the detected compact galaxy, we see no evidence of rotation or that the CO-emitting gas is spatially extended relative to the stellar light. This casts doubt on recent suggestions that the gas in these compact galaxies is rotating and significantly extended compared to the stars. Instead, we suggest that, at least for this object, the gas is centrally concentrated, and only traces a small fraction of the total galaxy dynamical mass.

\end{abstract}


%%%%%%%%%%%%%%%%%%%%%%%%%%%%%%%%%%%%%%%%%%%%%%%%%%%%%%%%%%%%%%%%%%%%%%%%%%%%%%%%%%%%%
%%%%%%%%%%%%%%%%%%%%%%%%%%%%%%%%%% Introduction %%%%%%%%%%%%%%%%%%%%%%%%%%%%%%%%%%%%%
%%%%%%%%%%%%%%%%%%%%%%%%%%%%%%%%%%%%%%%%%%%%%%%%%%%%%%%%%%%%%%%%%%%%%%%%%%%%%%%%%%%%%
\section{Introduction} \label{intro}

Galaxies with large stellar masses ($\Mstar \gtrsim 10^{11}$\,\msol) and little ongoing star formation have been observed at redshifts up to $z\sim4$ \citep{straatman14}, and begin to appear in large numbers by $z\sim2.5$ \citep[e.g.,][]{kriek06,whitaker10,cassata13}. Many studies have shown that these early quiescent galaxies were much smaller at $z\gtrsim1.5$ than equally massive star-forming galaxies at similar redshifts \citep[e.g.,][]{trujillo07,vanderwel14} or quiescent galaxies of similar mass in the local universe \citep[e.g.,][]{daddi05,trujillo06,buitrago08,cimatti08,vandokkum08,damjanov11}. With effective radii of just $\sim1-3$\,kpc, the stellar densities are of order 100$\times$ higher than present-day elliptical galaxies. Similarly massive and compact galaxies are extremely rare in the local universe \citep[e.g.,][]{trujillo09,taylor10}, implying significant size growth largely consistent with the effects of minor merging (e.g., \citealt{trujillo11,newman12}; see also \citealt{carollo13}). Although not all local massive galaxies had a compact progenitor \citep[e.g.,][]{franx08,vandokkum08}, most of the $z\sim2$ compact, massive galaxies likely now reside in the centers of present-day elliptical galaxies \citep[e.g.,][]{bezanson09,belli14a}.

The formation mechanism(s) of the $z\sim2$ compact quiescent population is still unclear.  Recently, however, a population of similarly compact yet highly star-forming galaxies at $z\sim2.5$ has been identified in deep \textit{Hubble Space Telescope} imaging \citep{barro13,barro14a,nelson14,williams14,vandokkum15}. Given their structural similarity, these compact star-forming galaxies (SFGs) are natural candidate progenitors of the early quiescent population, requiring only the cessation of star formation to superficially match the stellar distribution and structure of the $z\sim2$ quiescent galaxies. Additional evidence from dynamical studies \citep{barro14,vandokkum15} and number density evolution \citep{barro13} also indicates that compact SFGs will plausibly quench star formation on short timescales ($\lesssim$500\,Myr) to build up the growing quiescent population. The small sizes and non-exponential light profiles of compact SFGs present a clear contrast to the typical massive star-forming galaxies at $z\sim2$, which consist mostly of gas-rich, rapidly rotating disks \citep[e.g.,][]{wuyts11,tacconi13,wisnioski15}, evidence which suggests very different evolutionary histories.

The aforementioned dynamical studies indicate that the stellar masses of compact SFGs are ubiquitously comparable to or somewhat in excess of simple estimates of the total dynamical masses, \Mdyn, determined through observations of \halpha.  This implies that the dynamics of compact SFGs are almost completely dominated by the stars -- any gas present likely traces, but does not significantly contribute to, the gravitational potential. \citet{vandokkum15} argue that the \halpha-emitting gas is likely rotating and more extended than the stellar light, preventing the unphysical scenario of $\Mstar>\Mdyn$. This inference was motivated by observations of velocity gradients across the slit consistent with rotation for a few objects. Although selection effects are likely in play, these observations imply that the gas is extended by a factor $\sim$2.5 relative to the stars on average. Adaptive optics-assisted integral-field or interferometric synthesis imaging can be used to spatially and spectrally resolve the gas, providing a more robust tracer of the dynamics than can be inferred from long-slit spectroscopy.

Recent cosmological simulations have matched the observed number counts of compact SFGs and quiescent galaxies \citep{wellons15}. This work suggests that the early quiescent population consists mostly of a combination of galaxies which formed their stellar mass early and remained compact since their formation time, and objects which have recently undergone gas-rich major mergers. High-resolution hydrodynamical simulations further indicate that many characteristics of compact massive galaxies can be reproduced through in-situ formation through gas accretion and cooling \citep[e.g.,][]{naab09,feldmann15}, strong central star formation during a gas-rich major merger \citep[e.g.,][]{wuyts10,ceverino15}, and/or dissipative contraction of gas-rich disks \citep[e.g.,][]{ceverino15,zolotov15}. In each case, the properties of the gas are key, as its collisional nature allows energy to dissipate and angular momentum to be transferred through the galaxy, permitting the formation of characteristically dense stellar structures.

In this work, we present Karl G. Jansky Very Large Array (VLA) observations of the CO(1--0) line toward three compact SFGs at $z\sim2.3$. CO(1--0) has long been known as a tracer of molecular hydrogen in the interstellar medium (ISM), the direct fuel from which stars form. If, as we have outlined above, the compact SFGs are in the process of quenching star formation to become $z\sim2$ quiescent galaxies, we may expect molecular gas properties unlike those of normal SFGs at these redshifts. In particular, if compact SFGs are to quench in tandem with the rapid buildup of the quiescent population, we expect short gas depletion or quenching timescales, indicating that the ongoing high star formation rates (SFRs) cannot be sustained for more than a few hundred Myr. 

The outline of this paper is as follows. In Section~\ref{obs}, we describe our sample selection and VLA observations. Section~\ref{results} describes our results, including constraints on the molecular gas masses of our targeted objects (Section~\ref{mgas}), gas fractions and depletion timescales (Section~\ref{fgastdep}), and the physical extent of the molecular gas reservoirs in comparison to the stellar light (Section~\ref{dynamics}). We summarize our conclusions in Section~\ref{conclusions}.  Throughout, we assume a flat $\Lambda$CDM cosmology, with $H_0=67.7$\,\kms\,Mpc$^{-1}$ and $\Omega_\mathrm{m}=0.307$ \citep{planck15}.

%%%%%%%%%%%%%%%%%%%%%%%%%%%%%%%%%%%%%%%%%%%%%%%%%%%%%%%%%%%%%%%%%%%%%%%%%%%%%%%%%%%%%
%%%%%%%%%%%%%%%%%%%%%%%%%%%%%%%%%% Observations %%%%%%%%%%%%%%%%%%%%%%%%%%%%%%%%%%%%%
%%%%%%%%%%%%%%%%%%%%%%%%%%%%%%%%%%%%%%%%%%%%%%%%%%%%%%%%%%%%%%%%%%%%%%%%%%%%%%%%%%%%%
\section{Data} \label{obs}

%%%%%%%%%%%%%%%%%%%%%%%%%%%%%%%%%%% Selection %%%%%%%%%%%%%%%%%%%%%%%%%%%%%%%%%%%%%%%

\subsection{Selection of Compact Star-Forming Galaxies}

\begin{deluxetable*}{lllcccccc}
\tablecaption{Summary of VLA Observations \label{tab:obstable}}
\startdata
\tableline
\\
Source       & RA           & Dec                 & $z^\mathrm{a}$ & Array         & $\nu_{\mathrm{obs}}$ & Time On-Source & $\sigma^\mathrm{b}$ & Beam Size \\
             &              &                     &                & Configuration &  GHz                 &  h             & \uJy beam$^{-1}$  &        \\
\tableline
\vspace{-0.07in}
\\
\gn          & 12h36m27.73s & 62$^\circ$07'12.8'' & 2.301          & D             & 34.920                     & 1.7                & 82        & 2.3$\times$2.6'' \\
\ctt         & 10h00m17.15s &  2$^\circ$24'52.3'' & 2.469          & D             & 33.229                     & 5.1                & 64        & 2.6$\times$3.0'' \\
\cts         & 10h00m41.58s &  2$^\circ$27'51.5'' & 2.234          & C             & 35.644                     & 1.7                & 92        & 0.7$\times$0.8''        
\enddata
\tablenotetext{a}{Spectroscopic redshifts from \citet{vandokkum15}, determined from \halpha and \nii}
\tablenotetext{b}{rms noise in a 100\,\kms channel}
\end{deluxetable*}

\begin{figure}[hbt]
\includegraphics[width=\columnwidth]{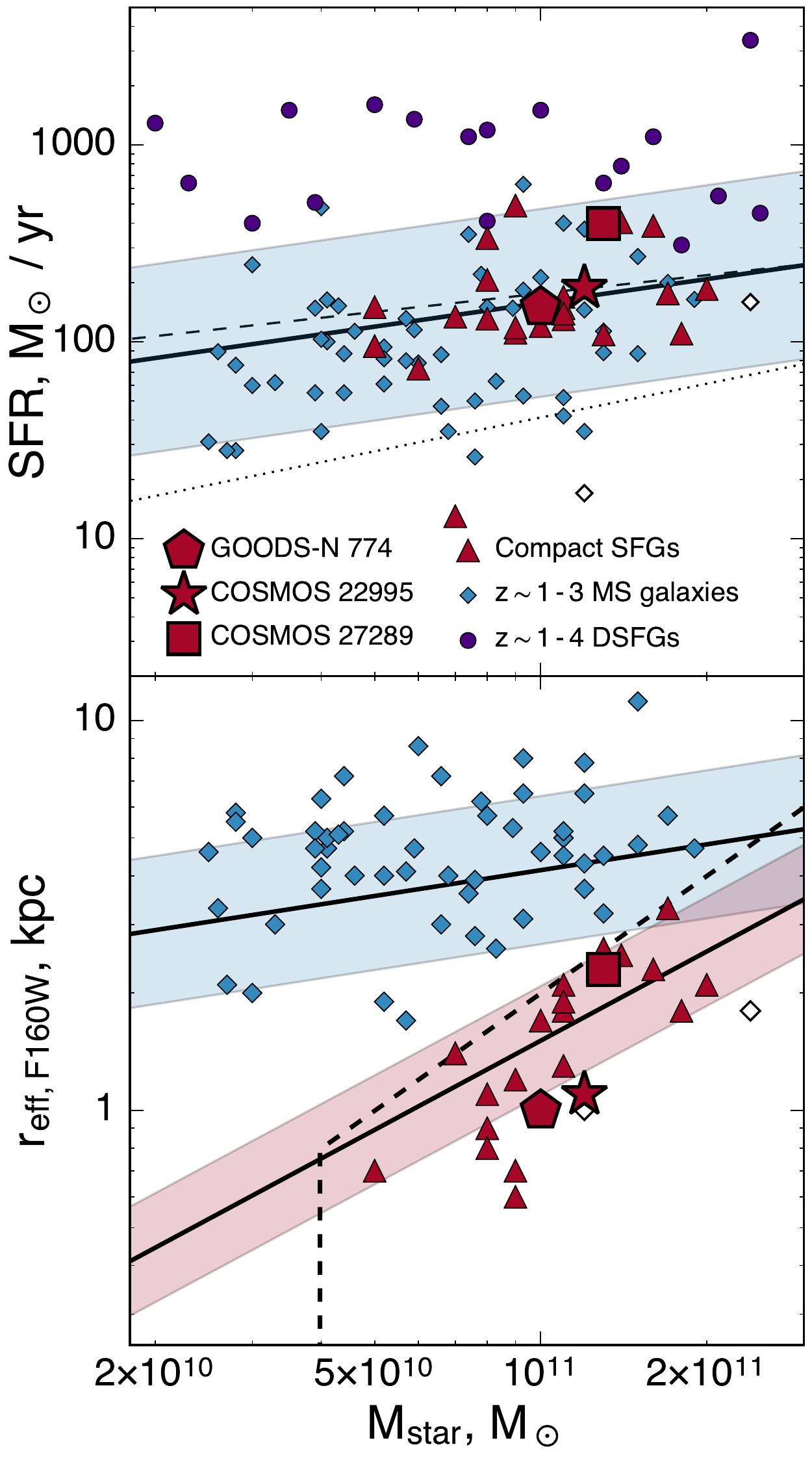}
\caption{
Selection of compact SFGs. In both panels, the full sample of \citet{vandokkum15} is shown as red triangles, while the objects observed by the VLA are shown with larger symbols and individually labeled. Comparison samples of CO-observed $z\sim1-3$ main sequence (MS) galaxies and $z\sim1-4$ dusty, star-forming galaxies (DSFGs) are shown as blue diamonds and navy circles, respectively; these samples are described further in Section~\ref{fgastdep}. Within the main sequence sample, two objects would also be selected as compact SFGs; these objects are shown with open symbols here and in Figures~\ref{fig:fgas} and \ref{fig:kstdep}.
\textit{Top:} The compact SFGs lie generally near the star-forming main sequence. The dotted, solid, and dashed black lines show the main sequence at redshifts 1, 2.3, and 3, respectively, as derived by \citet{whitaker12}, while the blue shaded region denotes SFRs a factor of 3 above and below the $z=2.3$ relation.  Note that the SFR of massive main sequence galaxies increases by approximately a factor of 4 over the redshift range spanned by the main-sequence comparison sample, $z\sim1-3$.
%\textit{Top:} The compact SFGs lie generally near the star-forming main sequence. All points are shown relative to the redshift-specific sequence derived by \citet{whitaker12}. The shaded region encompasses a factor of 4 above and below the main sequence, corresponding to its upper and lower range.
\textit{Bottom:} The compact SFGs have similar structural properties as quiescent galaxies at these redshifts.  The blue and red regions show the size--mass relations derived for star-forming and quiescent galaxies at $z\sim2.25$, respectively, by \citet{vanderwel14}. The size--mass selection criterion used by \citet{vandokkum15} is marked with a dashed black line.
}
\label{fig:selection}
\end{figure}

We selected compact SFGs for VLA CO(1--0) observations from the \citet{vandokkum15} sample; detailed selection criteria for the parent sample of compact SFGs are available in that work. In brief, the objects were selected from multi-band photometric catalogs in the CANDELS fields from the 3D-HST program \citep{koekemoer11,brammer12,skelton14,momcheva15} to be massive ($\Mstar>10^{10.6}$\,\msol), compact in WFC3 imaging \citep{vanderwel14}, and star-forming in rest-frame \textit{UVJ} color-color space. Catalog stellar masses were determined using stellar population synthesis modeling, and SFRs from UV+IR photometry assuming a Chabrier initial mass function, including \textit{Spitzer}/MIPS 24\,\um photometry to estimate the dust-obscured SFR of each object \citep{whitaker12,whitaker14}. These SFRs agree well with the 24\,\um-based calibrations derived by \citet{rujopakarn13}. Spectroscopic redshifts of each object were determined through Keck/MOSFIRE or NIRSPEC spectroscopy, detecting \halpha and \nii, and are accurate to $\Delta z \sim 0.001$. From the present sample, \ctt is identified as an X-ray AGN, while the other two sources are not.  We show the selection of compact SFGs in the SFR-\Mstar (the so-called star-forming main sequence) and $\mathrm{r}_{\mathrm{eff, F160W}}$-\Mstar planes in Figure~\ref{fig:selection}, highlighting the VLA-observed objects and including later comparison samples.

From this parent sample, we selected northern targets for VLA CO(1--0) observations, preferring objects with high apparent SFRs. The SFR of each target, ranging from $150-400$\,\msol/yr in the 3D-HST catalogs, implied that they likely would not be detected in \textit{Herschel} observations of the GOODS-N and COSMOS fields \citep{elbaz11,oliver12,roseboom12}. Indeed, only \cts was detected in any \textit{Herschel}/PACS or SPIRE bands\footnote{\citet{nelson14} report detections of \gn in all PACS and SPIRE bands, but we are unable to reproduce their photometry using 24\,\um cross-matched \textit{Herschel} catalogs \citep{roseboom12} or by visually inspecting the \textit{Herschel} images. The flux densities reported as significant detections by \citet{nelson14} are comparable to or fainter than the quoted \textit{Herschel} map depths. For consistency with the other sources and the 24\,\um cross-matched catalogs, we treat this object as undetected by \textit{Herschel}.}. A simple modified blackbody fit to the FIR photometry indicates SFR$= 250\pm80$\,\msol/yr for \cts, somewhat lower than its SFR in the 3D-HST catalogs. For the other two sources, the \textit{Herschel} non-detections indicate upper limits on the obscured SFR of $\lesssim 200$\,\msol/yr, assuming $\Tdust=30$\,K. These upper limits are consistent with the UV$+$24\,\um-derived catalog SFRs. For consistency in our own analysis and with \citet{vandokkum15}, we use the 3D-HST catalog UV$+$IR SFRs for all sources.

%%%%%%%%%%%%%%%%%%%%%%%%%%%%%%%%%% Observations %%%%%%%%%%%%%%%%%%%%%%%%%%%%%%%%%%%%%%

\subsection{VLA Observations}

VLA observations were carried out between October 2015 and March 2016 under programs 15B-283 and 16A-203 (PI J.~Spilker), and are summarized in Table~\ref{tab:obstable}. \gn and \ctt were observed in the VLA D-configuration (maximum baseline $\sim$1\,km), while \cts was observed in the C-configuration (maximum baseline $\sim$3.4\,km). Complex gain solutions were calculated using the quasars J1148+5924, J1008+062, and J0948+0022 for \gn, \ctt, and \cts, respectively, while observations of the quasar 3C286 were used both for bandpass and absolute flux calibration for all sources. The absolute flux scale is estimated to be accurate to 10\%. In all cases, the correlator was configured to deliver 2\,GHz of continuous bandwidth using the 8-bit samplers, with each baseband pair delivering eight dual-polarization basebands of 128\,MHz width and 1\,MHz channelization. The VLA Ka band receivers were tuned to center the CO(1--0) line in one of these basebands using the \halpha-based spectroscopic redshifts of \citet{vandokkum15}.

After calibration, the data were imaged using natural weighting, which maximizes point-source sensitivity. No source is detected in 9\,mm continuum emission, although we serendipitously detected a $S_\mathrm{35GHz}=170$\,\uJy continuum source in the \gn data located at 12h36m22.50s, +62$^\circ$06'53.9''. This source corresponds to GOODS-N 521 in the 3D-HST catalogs, and appears to be a $z\sim1.9$ galaxy. No lines are detected in this source, and we will not discuss it further.

\begin{figure}[htb]
\includegraphics[width=\columnwidth]{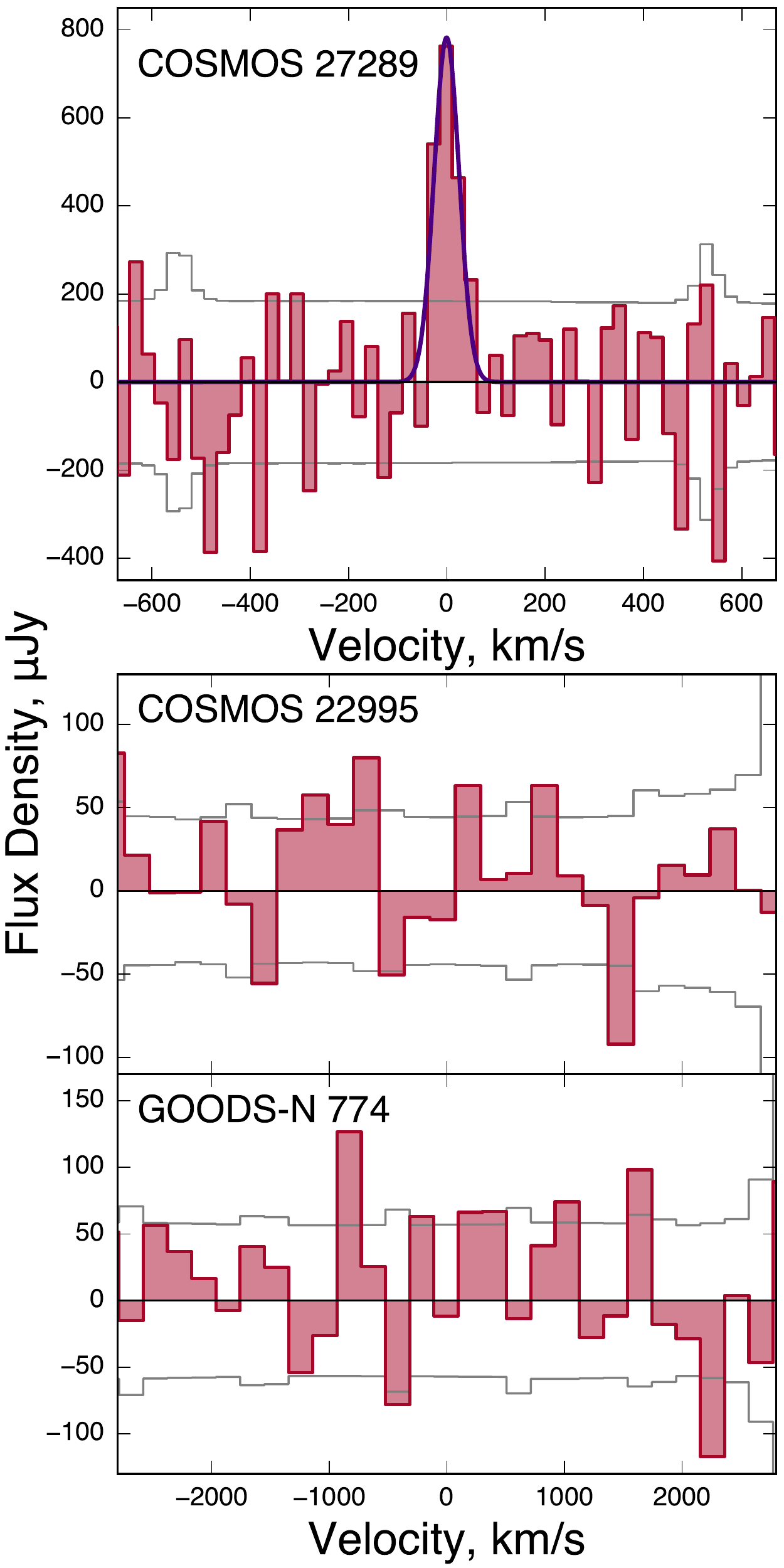}
\caption{
CO(1--0) spectra of the three sources observed in this work, derived by fitting a point source to the visibilities of each channel. For each source, we show the spectrum in red, with the frequency-dependent $\pm$1$\sigma$ noise as thin gray lines. For \cts, the spectrum is shown at 3\,MHz
($\sim$25\,\kms) resolution, and the best-fit Gaussian profile is shown in navy. For the other two sources, the spectra are plotted in 24\,MHz ($\sim$210\,\kms) channels.
}
\label{fig:spectra}
\end{figure}

Neither \gn or \ctt are detected in CO(1--0) emission. For both sources, we extract the spectra shown in Figure~\ref{fig:spectra} by fitting a point source to the visibilities averaged over 24\,MHz ($\sim$210\,\kms), fixing the position to the source position in \textit{HST} imaging, using the uncertainty from this procedure to estimate the noise in each channel. The noise is frequency-dependent due to the decline in sensitivity at the edges of each 128\,MHz baseband delivered by the VLA correlator. We use these spectra to place upper limits on the integrated CO flux assuming line widths of 500\,\kms, given in Table~\ref{tab:data}.

\begin{figure}[htb]
\includegraphics[width=\columnwidth]{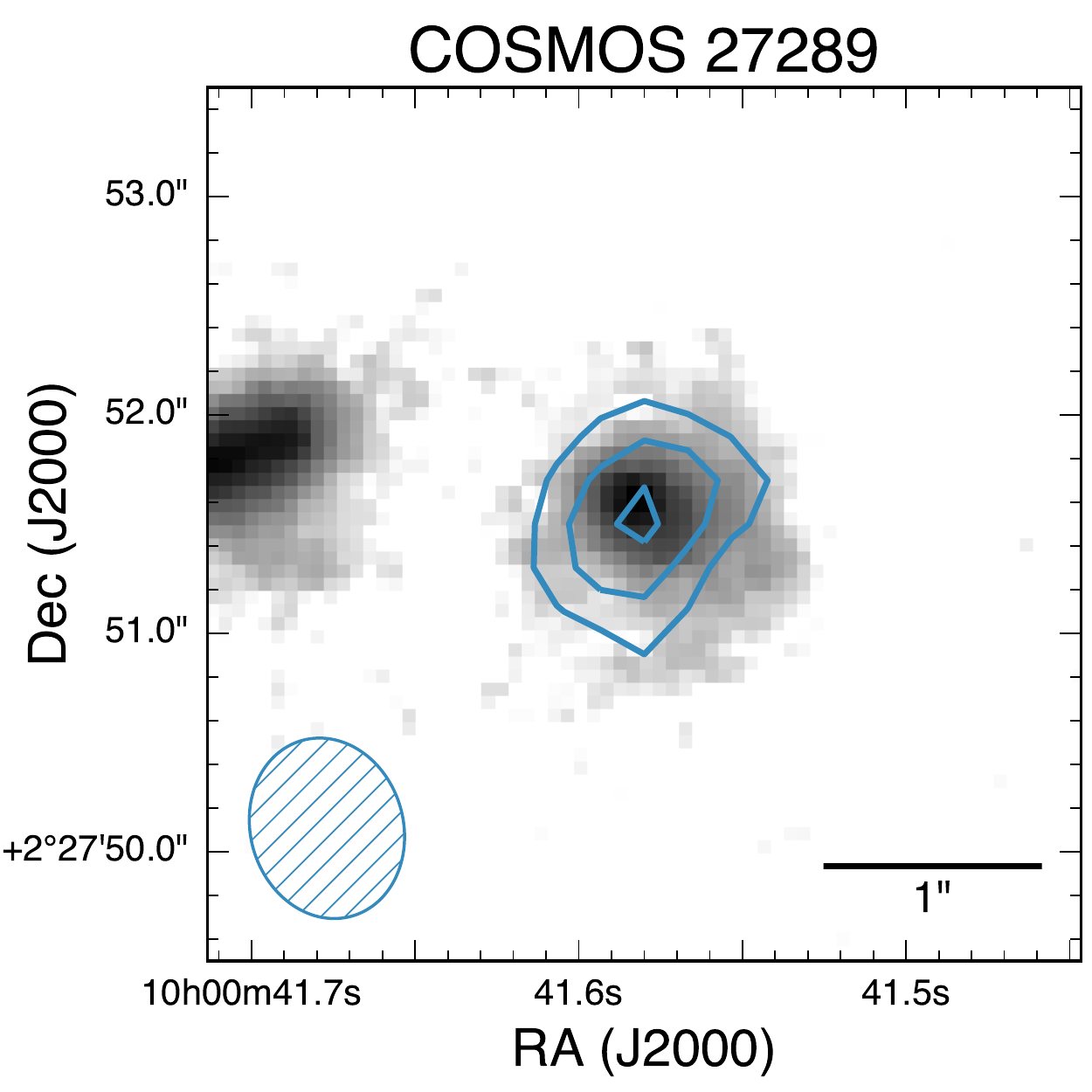}
\caption{
Contours of the CO(1--0) emission in \cts are overlaid on the \textit{HST}/F160W image of this object. The CO emission is averaged over 12\,MHz ($\sim$100\,\kms) centered on the line peak, and contours are drawn at 3, 5, and 7$\sigma$. The synthesized beam in this naturally-weighted image is shown in the lower left.
}
\label{fig:cdetect}
\end{figure}

\cts is significantly detected in CO(1--0) emission. A naturally-weighted integrated CO image of this galaxy is shown in Figure~\ref{fig:cdetect}. To determine if \cts is spatially resolved in the C-configuration data, we fit a point source to the visibilities averaged over 100\,\kms centered at the peak of the CO emission.  This velocity range encompasses $\sim$95\% of the integrated CO emission. No significant residual emission remains. We also examined the visibility amplitudes of the channels with significant line emission as a function of baseline length and found no significant decrease on long baselines, consistent with pointlike source structure. Finally, we created other images of the data applying various tapers in the $uv$-plane. In each case, we recover the same integrated flux density as we measured from the full data. These tests indicate that \cts is not extended at $\sim$0.75'' resolution, implying an upper limit on the CO size of $\mathrm{r}_{\mathrm{CO}}\lesssim 3$\,kpc.  We extract the spectrum shown in Figure~\ref{fig:spectra} by fitting a point source to the visibilities averaged over 3\,MHz ($\sim$25\,\kms), with the position fixed to the peak of the CO emission. We fit a simple Gaussian to this spectrum; the results are listed in Table~\ref{tab:data}.

\begin{deluxetable*}{lccccc}
\tablecaption{Observational Results \label{tab:data}}
\startdata
\tableline
\\
Source & \Mstar           & r$_\mathrm{eff, F160W}$ & SFR      & \lprime                     & FWHM$_\mathrm{CO}$ \\
       & $10^{11}$\,\msol & kpc                     & \msol/yr & $10^{10}$\,K\,\kms\,pc$^2$             & \kms               \\
\tableline
\vspace{-0.07in}
\\
\gn    & 1.0              & 1.0                     & 150      & $<1.4$         &  \\
\ctt   & 1.2              & 1.1                     & 190      & $<1.3$         &  \\
\cts   & 1.3              & 2.3                     & 400      & $1.3 \pm 0.3   $& 60$\pm$18 
\enddata
\tablecomments{Stellar masses are from the 3D-HST catalogs \citep{brammer12,skelton14},
source sizes from \citet{vanderwel14}, and SFRs from \citet{whitaker14}.
We list 3$\sigma$ \lprime
upper limits for the two non-detections, assuming a line width of 500\,\kms, 
approximately the median \halpha FWHM of
the \citet{vandokkum15} sample. These \lprime values are equivalent to \Mgas under the
assumption of $\alphaco=1$\,\msol (K\,\kms pc$^2$)$^{-1}$; see Section~\ref{mgas}.}
\end{deluxetable*}

%%%%%%%%%%%%%%%%%%%%%%%%%%%%%%%%%%%%%%%%%%%%%%%%%%%%%%%%%%%%%%%%%%%%%%%%%%%%%%%%%%%%%
%%%%%%%%%%%%%%%%%%%%%%%%%%%%%%%%%%%% Results %%%%%%%%%%%%%%%%%%%%%%%%%%%%%%%%%%%%%%%%
%%%%%%%%%%%%%%%%%%%%%%%%%%%%%%%%%%%%%%%%%%%%%%%%%%%%%%%%%%%%%%%%%%%%%%%%%%%%%%%%%%%%%
\section{Results and Discussion} \label{results}

%%%%%%%%%%%%%%%%%%%%%%%%%%%%%%%%%%%%%%%%%%%%%%%%%%%%%%%%%%%%%%%%%%%%%%%%%%%%%%%%%%%%%
%%%%%%%%%%%%%%%%%%%%%%%%%%%%%%%% Mgas, alphaCO %%%%%%%%%%%%%%%%%%%%%%%%%%%%%%%%%%%%%%
%%%%%%%%%%%%%%%%%%%%%%%%%%%%%%%%%%%%%%%%%%%%%%%%%%%%%%%%%%%%%%%%%%%%%%%%%%%%%%%%%%%%%
\subsection{Molecular Gas Masses and the CO-H$_2$ Conversion Factor} \label{mgas}

Central to our interpretation of the CO(1--0) observations is the conversion factor from CO luminosity to molecular gas mass, \alphaco. The CO-H$_2$ conversion factor is known to vary with the metallicity and kinematic state of the molecular gas (for a recent review, see \citealt{bolatto13}). For star-forming galaxies near solar metallicity, the value of \alphaco ranges from $\sim$0.8\,\msol (K\,\kms pc$^2$)$^{-1}$ in highly star-forming objects \citep[e.g.,][]{downes98} to $\sim$4\,\msol (K\,\kms pc$^2$)$^{-1}$ in the Milky Way and nearby quiescently star-forming galaxies (e.g., \citealt{sandstrom13}; hereafter we suppress the units of \alphaco).

The appropriate value of \alphaco in compact SFGs is not immediately obvious, but we can estimate its value and plausible upper and lower bounds through several methods. First, both the mass-metallicity relation and the \nii/\halpha line ratios of our sources indicate that the metallicity of each source is approximately solar or slightly super-solar \citep{pettini04,mannucci10}. In this regime, variations in \alphaco are no longer significantly affected by metallicity; instead, variations are driven by optical depth and/or excitation effects. This implies that \alphaco is almost certainly not significantly higher than the galactic value, $\alphaco \sim 4$. Several authors have derived theoretical or empirical formulations of the dependence of \alphaco on metallicity \citep[e.g.,][]{wolfire10,glover11,feldmann12} which also indicate that $\alphaco \sim4$ is a reasonable upper limit for galaxies of approximately solar metallicity.

We can place further limits on \alphaco by following the discussion of \citet{narayanan12}. These authors derived a fitting formula for \alphaco using hydrodynamical simulations of isolated and merging systems coupled with line and dust radiative transfer. This fitting formula depends on both the galaxy metallicity and the CO surface brightness, where objects with higher CO surface brightness have lower conversion factors. For the two non-detections, we assume that the extent of the CO(1--0) emission is at least as large as the stellar emission seen in \textit{HST}, which is nearly always observed \citep[e.g.,][]{tacconi13,spilker15}. This places an upper limit on the CO surface brightness.  At solar metallicity, our two non-detections imply conversion factors $\alphaco \gtrsim$0.7. For \cts, in contrast, our data place an upper limit on \alphaco. In this case, because the CO(1--0) emission is unresolved at 0.75'' resolution, the \citet{narayanan12} fitting formula implies $\alphaco \lesssim$1.6.

Finally, we can derive an empirical estimate of \alphaco by noting that it appears to correlate well with galaxy SFR or \lir \citep{spilker15}, based on a compilation of estimates of \alphaco in $z>1$ objects using various techniques from the literature, including dust-based, dynamics-based, and CO surface brightness-based methods. For the objects in our sample, this correlation implies $\alphaco\sim 1-3$, in reasonable agreement with our previous estimates.  

In summary, all indications are that the CO-H$_2$ conversion factor is expected to be relatively low in these highly star-forming objects. For simplicity we adopt $\alphaco=1$ for the remainder of this work. For \cts, this results in $\Mgas = (1.3 \pm 0.3) \times 10^{10}$\,\msol, and 3$\sigma$ upper limits of $\Mgas < 1.3 \times 10^{10}$ and $<1.4 \times 10^{10}$\,\msol for \ctt and \gn, respectively, integrating over a 500\,\kms line width (approximately the median \halpha FWHM of the \citealt{vandokkum15} sample).  We reiterate that these measurements carry significant systematic uncertainties, likely at least a factor of two. Future observations may help clarify the interpretation of CO(1--0) in compact SFGs by spatially and spectrally resolving the objects (yielding a dynamical constraint on \alphaco) or using observations of the long-wavelength dust continuum and dust-to-gas ratio arguments to derive alternate estimates of \Mgas.

%%%%%%%%%%%%%%%%%%%%%%%%%%%%%%%%%%%%%%%%%%%%%%%%%%%%%%%%%%%%%%%%%%%%%%%%%%%%%%%%%%%%%
%%%%%%%%%%%%%%%%%%%%%%%%%%%%%%%%%%%% fgas, tdep %%%%%%%%%%%%%%%%%%%%%%%%%%%%%%%%%%%%%
%%%%%%%%%%%%%%%%%%%%%%%%%%%%%%%%%%%%%%%%%%%%%%%%%%%%%%%%%%%%%%%%%%%%%%%%%%%%%%%%%%%%%
\subsection{Molecular Gas Fractions and Depletion Timescales} \label{fgastdep}

\begin{figure}[htb]
\includegraphics[width=\columnwidth]{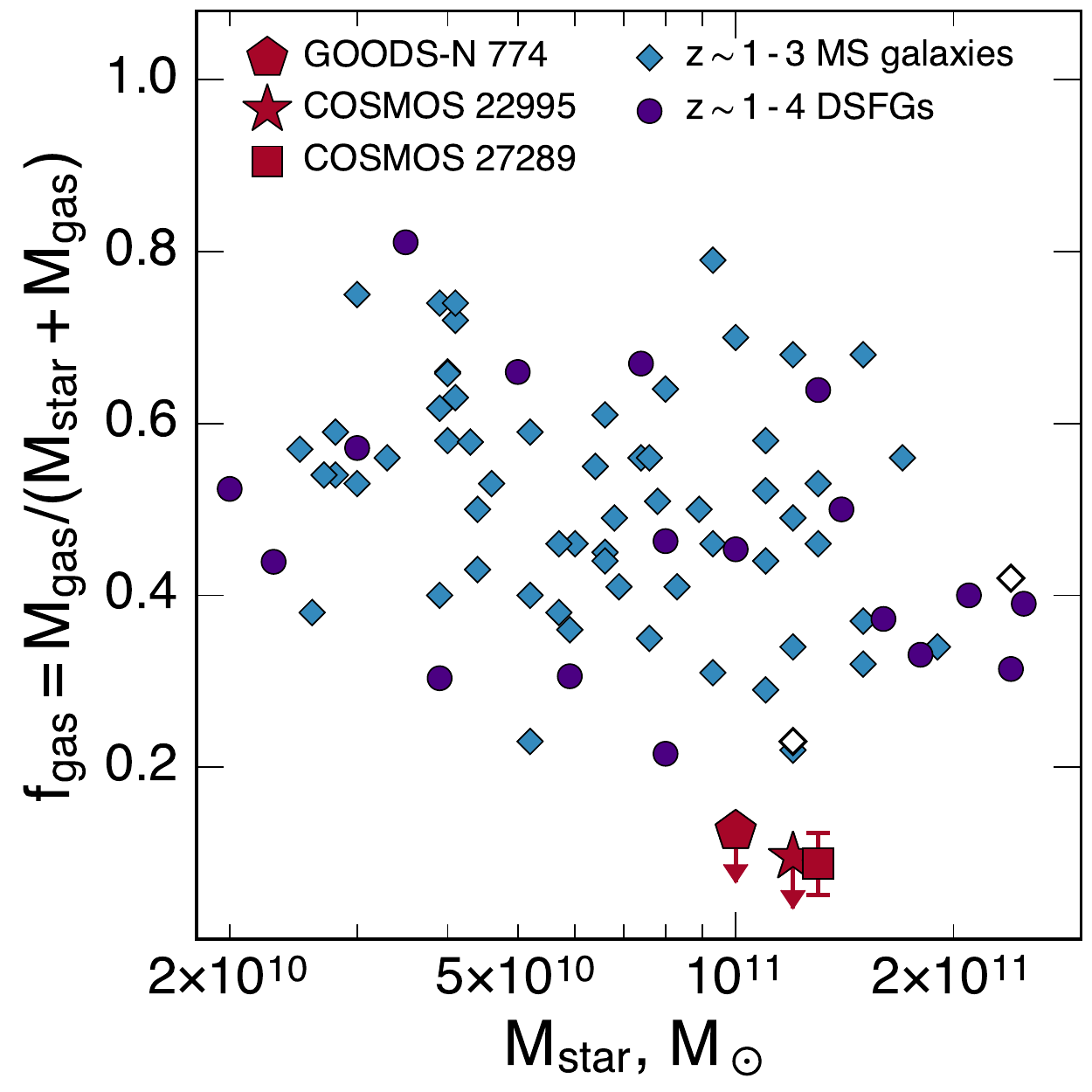}
\caption{
The gas fractions \fgas observed for the compact SFGs observed in this work (red symbols, assuming $\alphaco=1$; see Section~\ref{mgas}), compared to those measured for $z\sim1-3$ main-sequence (light blue diamonds) and starbursting galaxies (dark blue circles). The non-detections are 3$\sigma$ upper limits. The compact SFGs we have observed have very low gas fractions compared to any other objects at these redshifts with measured gas masses, a result which holds even if we adopt a Milky Way-like \alphaco. Open symbols are the two main sequence comparison objects which also meet the compactness selection criterion (Figure~\ref{fig:selection}).
}
\label{fig:fgas}
\end{figure}

Using the molecular gas masses derived in the previous section, in Figure~\ref{fig:fgas} we compare the baryonic gas fractions $\fgas \equiv \Mgas/(\Mstar+\Mgas)$ of the compact SFGs with those derived for both $z\sim1-3$ normal, main-sequence galaxies \citep{daddi10,tacconi13} and $z\sim1-4$ dusty, star-forming galaxies (DSFGs; compiled from \citealt{swinbank10,ivison11,ivison13,fu12,fu13,magnelli12,magdis12,hodge13,ma15,aravena16}). Gas masses for all objects were derived from CO (although largely from higher-excitation transitions instead of the ground state in the main-sequence sample), and all measurements are subject to significant systematic uncertainties. From this figure it is clear that the compact SFGs in our sample are severe outliers from other $z\sim2$ populations, even though the compact SFGs generally lie near the star-forming main sequence (Figure~\ref{fig:selection}; see also  \citealt{whitaker12,barro14}). With gas fractions $\lesssim$12\%, the compact SFGs are among the most gas-poor objects observed at $z\sim2$, in stark contrast to the high gas fractions ubiquitously observed at these redshifts (though see also \citealt{narayanan12b}). Even if we adopt a Milky Way-like $\alphaco=4$, the inferred baryonic gas fractions are $\lesssim$30\%, lower than 90\% of the CO-observed comparison sample objects.  In other words, the compact SFGs we have observed are genuinely deficient in CO emission, regardless of its conversion to \Mgas.

\begin{figure*}[htb]%
%\centering
\includegraphics[width=0.48\textwidth]{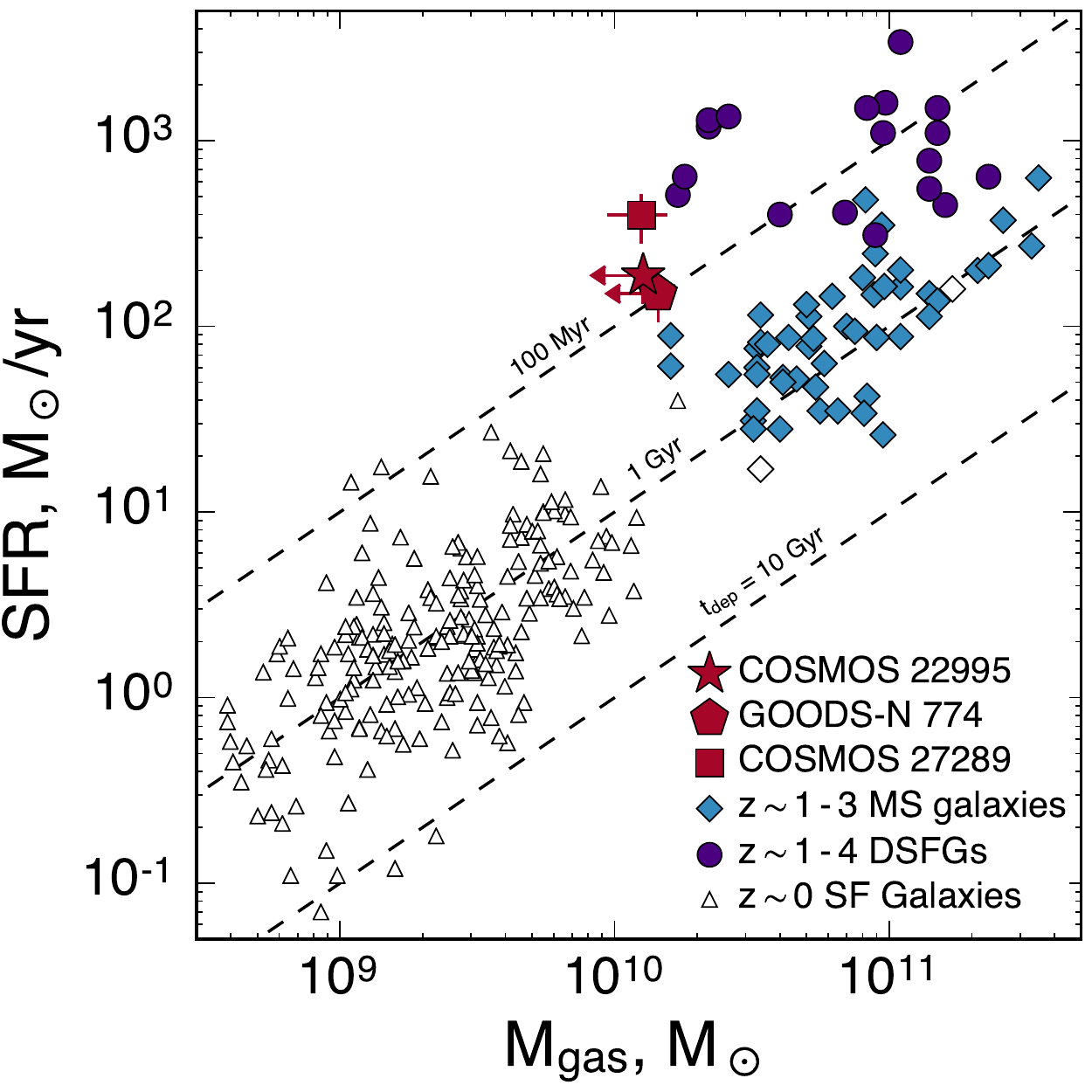}%
\includegraphics[width=0.48\textwidth]{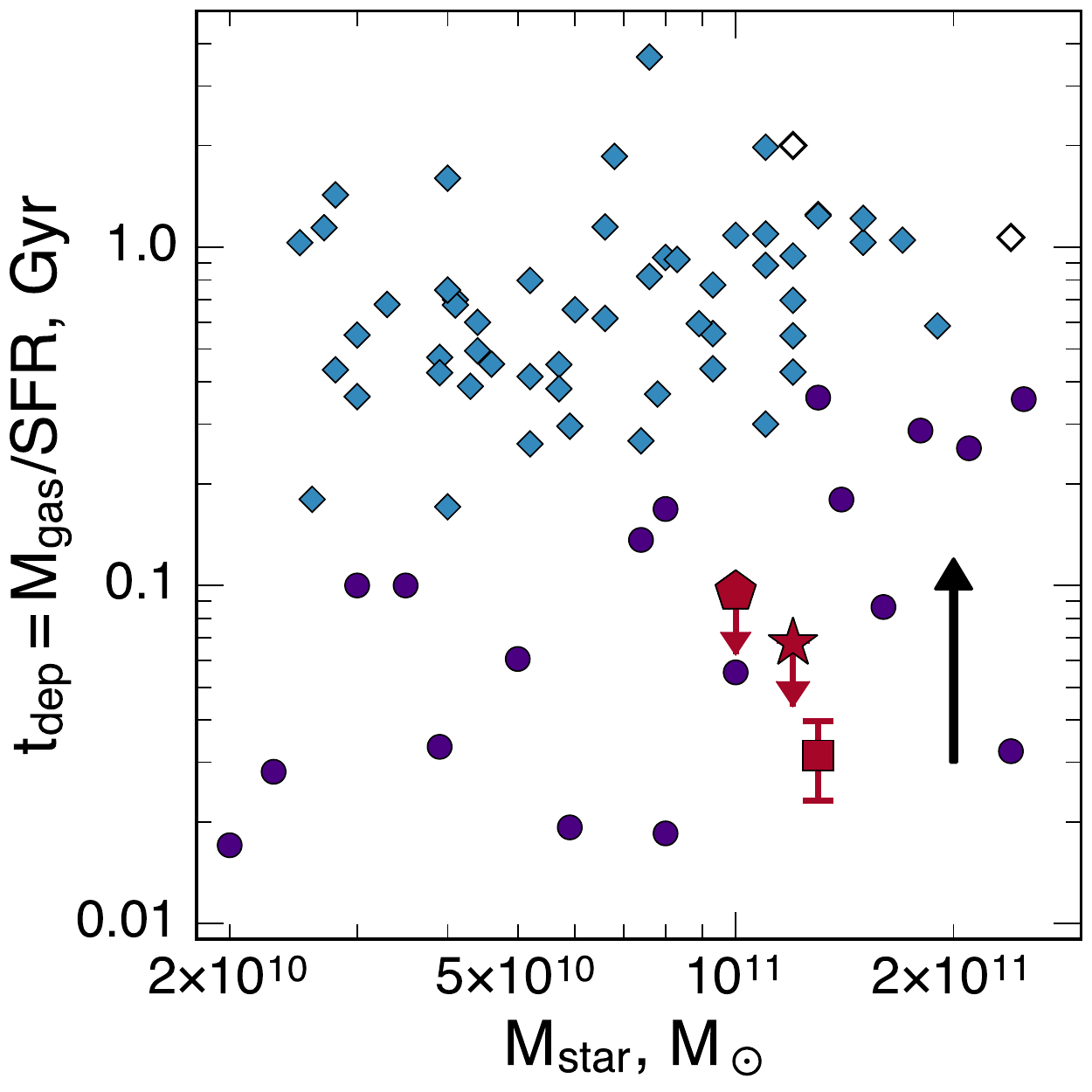}
\caption{
\textit{Left:} 
The integrated Schmidt-Kennicutt relation between SFR and \Mgas, with symbols as in Figure~\ref{fig:fgas}. We additionally include the $z\sim0$ sample of star-forming galaxies of \citealt{saintonge11} (white triangles), and show lines of constant \tdep (dashed).
\textit{Right:} 
The gas depletion times \tdep as a function of \Mstar, assuming the star formation in each object continues at its present rate. The compact SFGs we have observed show depletion times much shorter than typical galaxies at these redshifts, consistent with a rapid transition to quiescence. The arrow indicates the effect of adopting a galactic CO--H$_2$ conversion factor $\alphaco=4$.
}%
\label{fig:kstdep}%
\end{figure*}

A consequence of their low gas masses accompanied by rapid star formation is that the compact SFGs we have observed are similar outliers from main sequence populations in the Schmidt-Kennicutt relation between SFR and \Mgas and in their gas depletion timescales, $\tdep \equiv \Mgas/\mathrm{SFR}$, as shown in Figure~\ref{fig:kstdep}.
In this respect, the compact SFGs appear analogous to the highly star-forming DSFGs, with depletion times $\lesssim10^8$\,yr.
Whether the similar depletion times between compact SFGs and DSFGs is due to similar star-forming conditions in these objects is unclear. The low gas masses of compact SFGs indicate that they can undergo little further mass growth without replenishment of their gas reservoirs, in contrast to DSFGs, which have typical high gas fractions (Figure~\ref{fig:fgas}). The low gas fractions of compact SFGs seem to indicate that they are elevated relative to most objects on the Schmidt relation not because they exhibit very efficient star formation (high SFR at a given \Mgas), but because they are about to exhaust their gas reservoirs (low \Mgas at a given SFR).

Some authors have suggested an evolutionary connection between DSFGs and the $z\sim2$ compact quiescent population \citep[e.g.,][]{blain04,tacconi08,toft14} based in part on the observed small sizes of DSFGs \citep[e.g.,][]{simpson15b,spilker16}. The low gas masses of compact SFGs indicate that they may be transitioning below the main sequence, perhaps shortly after the end of the submillimeter-luminous phase. Only $\sim$15\% of compact SFGs are detected in \textit{Herschel}/SPIRE images \citep{barro14a}, consistent with this interpretation. Unfortunately, we have essentially no constraint on the past star formation in these objects, making it difficult to know if the compact SFGs were once DSFGs.

In both gas fraction and depletion time, the compact SFGs are outliers by factors of $\sim$5 to $>$10 from standard gas scaling relations \citep[e.g.,][]{genzel15}, indicating that these relations are not applicable to this unusual population of galaxies.  This result holds even if we adopt a galactic $\alphaco=4$, in which case compact SFGs still have gas fractions $\sim$2$\times$ lower and depletion times $\sim$3$\times$ lower than massive $z\sim2.3$ main sequence galaxies. Taken together, Figures~\ref{fig:fgas} and \ref{fig:kstdep} indicate that the compact SFGs we have selected are indeed consistent with being the immediate progenitors of some fraction of the $z\sim2$ quiescent population, in a state of rapid transition. If the current SFR continues at its present rate with no further gas accretion, each object we have observed will deplete its gas supply before $z=2$, presumably dropping below the star-forming main sequence at the same time. Our depletion time estimates could be lengthened if the SFR is declining and continues to decline as the gas supply is exhausted, but it is clear that compact SFGs can add very little additional stellar mass through star formation without substantial gas accretion. 

The low measured gas fractions and short depletion times are consistent with the inferences made by \citet{barro14}, who noted that simple dynamical mass estimates are frequently close to or lower than the stellar masses inferred from population synthesis models, implying gas fractions $\fgas \lesssim$30\%. A similar conclusion was inferred by \citet{vandokkum15}, who estimated gas fractions $\fgas \lesssim$40\%. 
Our results indicate even smaller gas fractions than estimated by these authors, affirming the conclusion that these galaxies have likely nearly exhausted their gas reservoirs and are either currently quenching or about to quench. The details of this quenching process are testable using higher spatial resolution observations. Observations of the dust continuum and high-resolution CO imaging would indicate whether the currently ongoing SFR and the reservoir of remaining gas are concentrated in the galaxy outskirts or the central regions. Such observations would provide a relatively straightforward test of theoretical models, which generally predict inside-out quenching \citep[e.g.,][]{zolotov15}.

%%%%%%%%%%%%%%%%%%%%%%%%%%%%%%%%%%%%%%%%%%%%%%%%%%%%%%%%%%%%%%%%%%%%%%%%%%%%%%%%%%%%%
%%%%%%%%%%%%%%%%%%%%%%%%%%%%%%%%%%%%%% Dynamics %%%%%%%%%%%%%%%%%%%%%%%%%%%%%%%%%%%%%
%%%%%%%%%%%%%%%%%%%%%%%%%%%%%%%%%%%%%%%%%%%%%%%%%%%%%%%%%%%%%%%%%%%%%%%%%%%%%%%%%%%%%
\subsection{A Compact Gas Reservoir in \cts} \label{dynamics}

We significantly detected CO(1--0) emission in \cts, finding a line width FWHM of $60\pm18$\,\kms. The \halpha line observed by \citet{vandokkum15} was somewhat broader, $130\pm30$\,\kms (though note this line was only marginally spectrally resolved; the NIRSPEC instrumental resolution at this wavelength is $\sim80$\,\kms). If real, the difference in line width between CO(1--0) and \halpha may indicate that the CO-emitting gas is slightly more compact than the \halpha-emitting gas, or that the \halpha line is somewhat contaminated by nuclear emission. Our C array observations did not spatially resolve the source in CO. This implies an upper limit on the CO-emitting region of $\mathrm{r}_{\mathrm{CO}}\lesssim 3$\,kpc.  Our ability to distinguish velocity gradients is somewhat hampered by low signal-to-noise, but we similarly see no strong evidence for a gradient that would be consistent with a rotating disk or any other organized velocity structure. Across the line profile, the emission centroid changes position by $\lesssim$0.8'' ($\lesssim$7\,kpc). In other words, our data offer no evidence that the CO-emitting gas is extended relative to the stellar emission ($\mathrm{r}_{\mathrm{eff, F160W}}=2.3$\,kpc), and only weak evidence of different spatial distributions of the CO- and \halpha-emitting gas. 

At least in this single case, our result contrasts with conclusion of \citet{vandokkum15}, who inferred that compact SFGs host rotating gas disks which are more extended than the stellar distributions by an average factor of $\sim$2.3.  This inference was motivated by noting that compact SFG gas velocity dispersions are systematically lower than expected from their stellar masses and sizes (equivalently, that stellar masses are higher than simple dynamical mass estimates), and supported by the observation that some objects showed \halpha velocity gradients across the slit consistent with rotation. For \cts, the gas radius that reconciles the stellar and dynamical masses is $\sim$20\,kpc \citep{vandokkum15}. At this large radius, the rotation curve implied by the stellar mass has fallen to match the low velocity measurement, but at the price of assuming that dark matter is negligible within 20\,kpc. However, such a large gas radius would have been easily detectable in our data. For this object, invoking a large, rotating gas disk is not supported by the data. 

How, then, can the discrepant dynamical and stellar mass estimates be reconciled? A simple, but unsatisfying explanation is that \cts is an outlier in terms of kinematic geometry or stellar mass estimate. \Mdyn depends on the assumed galaxy geometry and dynamics, resulting in factor of 2 systematic uncertainties. \cts may be an outlier compared to the assumed-disk-like overall population, or it may be that compact SFGs are not well-described by simple disk- or dispersion-dominated geometries (if, for example, they are generally recent merger remnants; \citealt{wellons15}).  If \cts does have disk-like dynamics, it may be face-on, although its axis ratio in the rest-frame optical implies an inclination of $\sim$45$^\circ$ \citep{vandokkum15}.
Second, we cannot rule out the possibility that stellar masses have been systematically overestimated in compact SFGs (but are accurate for most normal galaxies). If this is the case, the tension with the dynamical masses would be lowered while also bringing the gas fractions nearer to the value expected for normal, extended star-forming galaxies at this epoch. Given the extensive photometry and grism spectroscopy available for the extragalactic legacy fields in which compact SFGs have been selected, however, this seems unlikely.

A more likely solution is that the CO-emitting gas does not trace the full galaxy dynamical (or stellar) mass.  The calculations of \citet{vandokkum15} required a very extended gas disk because of their assumption that the gas velocity dispersion must trace the full stellar mass of the compact SFGs -- for a given dynamical model, if the total mass and circular velocity are fixed, the only remaining free parameter is the galaxy radius. Instead, our data paint a different picture.  Given that the observed gas radius and velocity dispersion are small, the gas likely only traces a small fraction of the total galaxy mass in the central regions. In other words, instead of fixing the total mass and velocity dispersion to infer large radii, we argue that the mass interior to the gas effective radius is much lower than the full stellar mass, removing the need to invoke large gas disks to reconcile the stellar and dynamical masses.  

In addition to our own data, this scenario is supported by the recent observations of \citet{barro16}, who observed that the stellar absorption lines were nearly 70\% broader than the gas nebular emission lines in a compact SFG at $z=1.7$. This difference increases the dynamical mass inferred from the stars by nearly a factor of 3, and prevents the unphysical scenario of $\Mstar > \Mdyn$. A similar effect is likely at work in \cts, predicting that future ultra-deep NIR spectroscopy will find a stellar velocity dispersion in excess of the gas velocity dispersion. More immediate progress may be made by continued CO observations using either higher-resolution B array observations or simply deeper observations with the current C array. 

Having only detected and placed constraints on the dynamics of a single object, we cannot draw strong conclusions concerning the discrepancy between stellar and dynamical masses in the compact SFG population as a whole. While larger samples of objects with spatially and spectrally resolved spectroscopy will be required to investigate this issue further, it is clear that even relatively modest upper limits on the spatial extent of the molecular gas provide powerful constraints on the gas dynamics of compact SFGs.

%%%%%%%%%%%%%%%%%%%%%%%%%%%%%%%%%%%%%%%%%%%%%%%%%%%%%%%%%%%%%%%%%%%%%%%%%%%%%%%%%%%%%
%%%%%%%%%%%%%%%%%%%%%%%%%%%%%%%%%% Conclusions %%%%%%%%%%%%%%%%%%%%%%%%%%%%%%%%%%%%%%
%%%%%%%%%%%%%%%%%%%%%%%%%%%%%%%%%%%%%%%%%%%%%%%%%%%%%%%%%%%%%%%%%%%%%%%%%%%%%%%%%%%%%
\section{Conclusions} \label{conclusions}

We have observed the CO(1--0) transition in three $z\sim2.3$ compact star-forming galaxies in an effort to determine the amount of molecular gas in these objects. Our two non-detections and one detected source confirm low baryonic gas fractions and very short gas depletion timescales, as expected if compact SFGs are currently or on the verge of quenching star formation. The gas fractions and depletion times are much lower than predicted from gas scaling relations. In the detected galaxy, we see no evidence that the CO-emitting gas is spatially extended relative to the stellar light, and only weak evidence for a difference in the CO dynamics compared to the \halpha emission. This result demonstrates that invoking large gas disks to increase dynamical masses estimated from gas kinematics above the stellar masses may not be warranted. Instead, we argue that the gas is centrally concentrated, and therefore need not kinematically trace the full stellar mass of compact SFGs. An increased sample size and high-resolution follow-up observations of the molecular gas content of compact SFGs would indicate whether this is true of the bulk of the population or if \cts is an outlier.

\acknowledgements{
J.S.S. and D.P.M. acknowledge support from the U.S. National Science Foundation under grant No. AST-1312950.
R.B. and K.E.W. gratefully acknowledge support by NASA through Hubble Fellowship grants \#HF-51318 and \#HF-51368 awarded by the Space Telescope Science Institute, which is operated by the Association of Universities for Research in Astronomy, Inc., for NASA, under contract NAS5-26555.
C.C.W. was supported by the NIRCam contract to the University of Arizona, NAS5-02015.
The National Radio Astronomy Observatory is a facility of the National Science Foundation operated under cooperative agreement by Associated Universities, Inc.
This work is based on observations taken by the 3D-HST Treasury Program (GO 12177 and 12328) with the NASA/ESA HST, which is operated by the Association of Universities for Research in Astronomy, Inc., under NASA contract NAS5-26555.
This research has made use of NASA's Astrophysics Data System.
}

{\it Facility:} \facility{VLA}

\bibliographystyle{apj}
%\bibliography{../spt_js_merge.bib}

\end{document}